\documentclass[,final,numberedheadings]{aipproc}

\layoutstyle{6x9}

\usepackage{amsmath}
\usepackage{amsfonts}
\usepackage{amssymb}
\usepackage{graphicx}

\input{tcilatex}

\begin{document}

\title{An Application of Reversible Entropic Dynamics on Curved Statistical
Manifolds\thanks%
{Presented at MaxEnt 2006, the 26th International Workshop on Bayesian Inference and
Maximum Entropy Methods (July 8-13, 2006, Paris, France)
}}

\classification{}
\keywords{Inductive inference, information geometry, statistical manifolds, relative entropy.
}

\author{Carlo Cafaro\thanks{%
E-mail: carlocafaro2000@yahoo.it}, \ S. A. Ali\thanks{%
E-mail: alis@alum.rpi.edu} \ and Adom Giffin\thanks{%
E-mail: physics101@gmail.com}}{
address={Department of Physics, University at Albany--SUNY, Albany, NY 12222, USA}}

\begin{abstract}
Entropic Dynamics (ED) $\left[ 1\right] $ is a theoretical framework
developed to investigate the possibility that laws of physics reflect laws
of inference rather than laws of nature. In this work, a RED (Reversible
Entropic Dynamics) model is considered. The geometric structure underlying
the curved statistical manifold $\mathcal{M}_{s}$ is studied. The
trajectories of this particular model are hyperbolic curves (geodesics) on $%
\mathcal{M}_{s}$. Finally, some analysis concerning the stability of these
geodesics on $\mathcal{M}_{s}$ is carried out.
\end{abstract}

\maketitle

\section{Introduction}

We use Maximum relative Entropy (ME) methods $[2,3]$ to construct a RED
model. ME methods are inductive inference tools. They are used for updating
from a prior to a posterior distribution when new information in the form of
constraints becomes available. We use known techniques $\left[ 1\right] $ to
show that they lead to equations that are analogous to equations of motion.
Information is processed using ME methods in the framework of Information
Geometry (IG) $\left[ 4\right] $. The ED model follows from an assumption
about what information is relevant to predict the evolution of the system.
We focus only on reversible aspects of the ED model. In this case, given a
known initial state and that the system evolves to a final known state, we
investigate the possible trajectories of the system. Reversible and
irreversible aspects in addition to further developments on the ED model are
presented in a forthcoming paper $\left[ 5\right] $. Given two probability
distributions, how can one define a notion of "distance" between them? The
answer to this question is provided by IG. Information Geometry is
Riemannian geometry applied to probability theory. As it is shown in $[6$, $%
7]$, the notion of distance between dissimilar probability distributions is
quantified by the Fisher-Rao information metric tensor.

\section{The RED Model}

We consider a RED model whose microstates span a $2D$ space labelled by the
variables $x_{1}\in 
\mathbb{R}
^{+}$ and $x_{2}\in 
\mathbb{R}
$. We assume the only testable information pertaining to the quantities $%
x_{1}$ and $x_{2}$ consists of the expectation values $\left\langle
x_{1}\rangle \text{, }\langle x_{2}\right\rangle $ and the variance $\Delta
x_{2}$. These three expected values define the $3D$ space of macrostates of
the system. Our model may be extended to more elaborate systems where higher
dimensions are considered. However, for the sake of clarity, we restrict our
consideration to this relatively simple case. A measure of
distinguishability among the states of the ED model is achieved by assigning
a probability distribution $p^{(tot)}\left( \vec{x}|\vec{\theta}\right) $ to
each macrostate $\vec{\theta}$ . The process of assigning a probability
distribution to each state provides $\mathcal{M}_{S}$ with a metric
structure. Specifically, the Fisher-Rao information metric defined in $(6)$
is a measure of distinguishability among macrostates. It assigns an IG to
the space of states.

\subsection{The Statistical Manifold $\mathcal{M}_{S}$}

Consider a hypothetical physical system evolving over a two-dimensional
space.\ The variables $x_{1}$ and $x_{2}$ label the $2D$ space of
microstates of the system. We assume that all information relevant to the
dynamical evolution of the system is contained in the probability
distributions. For this reason, no other information is required. Each
macrostate may be thought as a point of a three-dimensional statistical
manifold with coordinates given by the numerical values of the expectations $%
\theta _{1}^{\left( 1\right) }=\left\langle x_{1}\right\rangle $, $\theta
_{1}^{\left( 2\right) }=\left\langle x_{2}\right\rangle $, $\theta
_{2}^{\left( 2\right) }=\Delta x_{2}$. The available information can be
written in the form of the following constraint equations,%
\begin{equation}
\begin{array}{c}
\left\langle x_{1}\right\rangle =\int_{0}^{+\infty }dx_{1}x_{1}p_{1}\left(
x_{1}|\theta _{1}^{\left( 1\right) }\right) \text{, }\left\langle
x_{2}\right\rangle =\int_{-\infty }^{+\infty }dx_{2}x_{2}p_{2}\left(
x_{2}|\theta _{1}^{\left( 2\right) },\theta _{2}^{\left( 2\right) }\right) 
\text{,} \\ 
\\ 
\Delta x_{2}=\sqrt{\left\langle \left( x_{2}-\left\langle x_{2}\right\rangle
\right) ^{2}\right\rangle }=\left[ \int_{-\infty }^{+\infty }dx_{2}\left(
x_{2}-\left\langle x_{2}\right\rangle \right) ^{2}p_{2}\left( x_{2}|\theta
_{1}^{\left( 2\right) },\theta _{2}^{\left( 2\right) }\right) \right] ^{%
\frac{1}{2}}\text{,}%
\end{array}%
\end{equation}%
where $\theta _{1}^{\left( 1\right) }=\left\langle x_{1}\right\rangle $, $%
\theta _{1}^{\left( 2\right) }=\left\langle x_{2}\right\rangle $ and $\theta
_{2}^{\left( 2\right) }=\Delta x_{2}$. The probability distributions $p_{1}$
and $p_{2}$ are constrained by the conditions of normalization,%
\begin{equation}
\int_{0}^{+\infty }dx_{1}p_{1}\left( x_{1}|\theta _{1}^{\left( 1\right)
}\right) =1\text{, }\int_{-\infty }^{+\infty }dx_{2}p_{2}\left( x_{2}|\theta
_{1}^{\left( 2\right) },\theta _{2}^{\left( 2\right) }\right) =1\text{.}
\end{equation}%
Information theory identifies the exponential distribution as the maximum
entropy distribution if only the expectation value is known. The Gaussian
distribution is identified as the maximum entropy distribution if only the
expectation value and the variance are known. ME methods allow us to
associate a probability distribution $p^{(tot)}\left( \vec{x}|\vec{\theta}%
\right) $ to each point in the space of states $\vec{\theta}\equiv \left(
\theta _{1}^{\left( 1\right) }\text{, }\theta _{1}^{\left( 2\right) }\text{, 
}\theta _{2}^{\left( 2\right) }\right) $. The distribution that best
reflects the information contained in the prior distribution $m\left( \vec{x}%
\right) $ updated by the information $(\left\langle x_{1}\right\rangle
,\left\langle x_{2}\right\rangle ,\Delta x_{2})$ is obtained by maximizing
the relative entropy 
\begin{equation}
S\left( \vec{\theta}\right) =-\int_{0}^{+\infty }\int_{-\infty }^{+\infty
}dx_{1}dx_{2}p^{(tot)}\left( \vec{x}|\vec{\theta}\right) \log \left( \frac{%
p^{(tot)}\left( \vec{x}|\vec{\theta}\right) }{m\left( \vec{x}\right) }%
\right) \text{,}
\end{equation}%
where $m(\vec{x})\equiv m$ is the uniform prior probability distribution.
The prior $m\left( \vec{x}\right) $ is set to be uniform since we assume the
lack of prior available information about the system (postulate of equal 
\textit{a priori} probabilities). Upon maximizing $\left( 3\right) $, given
the constraints $\left( 1\right) $ and $\left( 2\right) $, we obtain%
\begin{equation}
p^{(tot)}\left( \vec{x}|\vec{\theta}\right) =p_{1}\left( x_{1}|\theta
_{1}^{\left( 1\right) }\right) p_{2}\left( x_{2}|\theta _{1}^{\left(
2\right) },\theta _{2}^{\left( 2\right) }\right) =\frac{1}{\mu _{1}}e^{-%
\frac{x_{1}}{\mu _{1}}}\frac{1}{\sqrt{2\pi \sigma _{2}^{2}}}e^{-\frac{%
(x_{2}-\mu _{2})^{2}}{2\sigma _{2}^{2}}},
\end{equation}%
where $\theta _{1}^{\left( 1\right) }=\mu _{1}$, $\theta _{1}^{\left(
2\right) }=\mu _{2}$ and $\theta _{2}^{\left( 2\right) }=\sigma _{2}$. The
probability distribution $(4)$ encodes the available information concerning
the system. Note that we have assumed uncoupled constraints between the
microvariables $x_{1}$ and $x_{2}$. In other words, we assumed that
information about correlations between the microvariables need not to be
tracked. This assumption leads to the simplified product rule $\left(
4\right) $. Coupled constraints however, would lead to a generalized product
rule in $\left( 4\right) $ and to a metric tensor $\left( 7\right) $ with
non-trivial off-diagonal elements (covariance terms). Correlation terms may
be fictitious. They may arise for instance from coordinate transformations.
On the other hand, correlations may arise from external fields in which the
system is immersed. In such situations, correlations between $x_{1}$\ and $%
x_{2}$ effectively describe interaction between the microvariables and the
external fields. Such generalizations would require more delicate analysis.

\section{The Metric Structure of $\mathcal{M}_{s}$}

We cannot determine the evolution of microstates of the system since the
available information is insufficient. Not only is the information available
insufficient but we also do not know the equation of motion. In fact there
is no standard "equation of motion".\ Instead we can ask: how close are the
two total distributions with parameters $(\mu _{1},\mu _{2},\sigma _{2})$
and $(\mu _{1}+d\mu _{1},\mu _{2}+d\mu _{2},\sigma _{2}+d\sigma _{2})$? Once
the states of the system have been defined, the next step concerns the
problem of quantifying the notion of change from the state $\vec{\theta}$ to
the state $\vec{\theta}+d\vec{\theta}$. A convenient measure of change is
distance. The measure we seek is given by the dimensionless "distance" $ds$
between $p^{(tot)}\left( \vec{x}|\vec{\theta}\right) $ and $p^{(tot)}\left( 
\vec{x}|\vec{\theta}+d\vec{\theta}\right) $ $\left[ 4\right] :$%
\begin{equation}
ds^{2}=g_{ij}d\theta ^{i}d\theta ^{j}\text{,}
\end{equation}%
where%
\begin{equation}
g_{ij}=\int d\vec{x}p^{(tot)}\left( \vec{x}|\vec{\theta}\right) \frac{%
\partial \log p^{(tot)}\left( \vec{x}|\vec{\theta}\right) }{\partial \theta
^{i}}\frac{\partial \log p^{(tot)}\left( \vec{x}|\vec{\theta}\right) }{%
\partial \theta ^{j}}
\end{equation}%
is the Fisher-Rao metric $\left[ 6\text{, }7\right] $. Substituting $\left(
4\right) $ into $\left( 6\right) $, the metric $g_{ij}$ on $\mathcal{M}_{s}$
becomes,%
\begin{equation}
g_{ij}=\left( 
\begin{array}{ccc}
\frac{1}{\mu _{1}^{2}} & 0 & 0 \\ 
0 & \frac{1}{\sigma _{2}^{2}} & 0 \\ 
0 & 0 & \frac{2}{\sigma _{2}^{2}}%
\end{array}%
\right) \text{.}
\end{equation}%
From $\left( 7\right) $, the "length" element $\left( 5\right) $ reads,%
\begin{equation}
ds^{2}=\frac{1}{\mu _{1}^{2}}d\mu _{1}^{2}+\frac{1}{\sigma _{2}^{2}}d\mu
_{2}^{2}+\frac{2}{\sigma _{2}^{2}}d\sigma _{2}^{2}\text{.}
\end{equation}%
We bring attention to the fact that the metric structure of $\mathcal{M}_{s}$
is an emergent (not fundamental) structure. It arises only after assigning a
probability distribution $p^{(tot)}\left( \vec{x}|\vec{\theta}\right) $ to
each state $\vec{\theta}$.

\subsection{The Statistical Curvature of $\mathcal{M}_{s}$}

We study the curvature of $\mathcal{M}_{s}$. This is achieved via
application of differential geometry methods to the space of probability
distributions. As we are interested specifically in the curvature properties
of $\mathcal{M}_{s}$, recall the definition of the Ricci scalar $R$,%
\begin{equation}
R=g^{ij}R_{ij}\text{,}
\end{equation}%
where $g^{ik}g_{kj}=\delta _{j}^{i}$ so that $g^{ij}=\left( g_{ij}\right)
^{-1}=diag(\mu _{1}^{2}$,$\sigma _{2}^{2}$,$\frac{\sigma _{2}^{2}}{2})$. The
Ricci tensor $R_{ij}$ is given by,%
\begin{equation}
R_{ij}=\partial _{k}\Gamma _{ij}^{k}-\partial _{j}\Gamma _{ik}^{k}+\Gamma
_{ij}^{k}\Gamma _{kn}^{n}-\Gamma _{ik}^{m}\Gamma _{jm}^{k}\text{.}
\end{equation}%
The Christoffel symbols $\Gamma _{ij}^{k}$ appearing in the Ricci tensor are
defined in the standard way, 
\begin{equation}
\Gamma _{ij}^{k}=\frac{1}{2}g^{km}\left( \partial _{i}g_{mj}+\partial
_{j}g_{im}-\partial _{m}g_{ij}\right) .
\end{equation}%
Using $\left( 7\right) $ and the definitions given above, the non-vanishing
Christoffel symbols are $\Gamma _{11}^{1}=-\frac{1}{\mu _{1}}$, $\Gamma
_{22}^{3}=\frac{1}{2\sigma _{2}}$, $\Gamma _{33}^{3}=-\frac{1}{\sigma _{2}}$
and $\Gamma _{23}^{2}=\Gamma _{32}^{2}=-\frac{1}{\sigma _{2}}$. The Ricci
scalar becomes%
\begin{equation}
R=-1<0\text{.}
\end{equation}%
From $(12)$ we conclude that $\mathcal{M}_{s}$ is a $3D$ curved manifold of
constant negative $(R=-1)$ curvature.

\section{Canonical Formalism for the RED Model}

We remark that RED can be derived from a standard principle of least action
(Maupertuis- Euler-Lagrange-Jacobi-type) $\left[ 1,8\right] $. The main
differences are that the dynamics being considered here, namely Entropic
Dynamics, is defined on a space of probability distributions $\mathcal{M}%
_{s} $, not on an ordinary vectorial space $V$ and the standard coordinates $%
q_{j} $ of the system are replaced by statistical macrovariables $\theta
^{j} $.

Given the initial macrostate and that the system evolves to a final
macrostate, we investigate the expected trajectory of the system on $%
\mathcal{M}_{s}$. It is known $\left[ 8\right] $ that the classical dynamics
of a particle can be derived from the principle of least action in the form,%
\begin{equation}
\delta J_{Jacobi}\left[ q\right] =\delta \int_{s_{i}}^{s_{f}}ds\mathcal{F}%
\left( q_{j},\frac{dq_{j}}{ds},s,H\right) =0\text{,}
\end{equation}%
where $q_{j}$ are the coordinates of the system, $s$ is an arbitrary
(unphysical) parameter along the trajectory. The functional $\mathcal{F}$
does not encode any information about the time dependence and it is defined
by,%
\begin{equation}
\mathcal{F}\left( q_{j},\frac{dq_{j}}{ds},s,H\right) \equiv \left[ 2\left(
H-U\right) \right] ^{\frac{1}{2}}\left( \underset{j,k}{\sum }a_{jk}\frac{%
dq_{j}}{ds}\frac{dq_{k}}{ds}\right) ^{\frac{1}{2}}\text{,}
\end{equation}%
where the energy of the particle is given by%
\begin{equation}
H\equiv E=T+U\left( q\right) =\frac{1}{2}\underset{j,k}{\sum }a_{jk}\left(
q\right) \dot{q}_{j}\dot{q}_{k}+U\left( q\right) \text{.}
\end{equation}%
The coefficients $a_{jk}\left( q\right) $ are the reduced mass matrix
coefficients and $\dot{q}=\frac{dq}{ds}$. We now seek the expected
trajectory of the system assuming it evolves from the given initial state $%
\theta _{old}^{\mu }=\theta ^{\mu }$\ $\equiv \left( \mu _{1}\left(
s_{i}\right) ,\mu _{2}\left( s_{i}\right) ,\sigma _{2}\left( s_{i}\right)
\right) $ to a new state $\theta _{new}^{\mu }=\theta ^{\mu }+d\theta ^{\mu
}\equiv \left( \mu _{1}\left( s_{f}\right) ,\mu _{2}\left( s_{f}\right)
,\sigma _{2}\left( s_{f}\right) \right) $. It can be shown that\ the system
moves along a geodesic in the space of states $\left[ 1\right] $.\ Since the
trajectory of the system is a geodesic, the RED-action is itself the length:%
\begin{equation}
J_{RED}\left[ \theta \right] =\int_{s_{i}}^{s_{f}}ds\left( g_{ij}\frac{%
d\theta ^{i}\left( s\right) }{ds}\frac{d\theta ^{j}\left( s\right) }{ds}%
\right) ^{\frac{1}{2}}\equiv \int_{s_{i}}^{s_{f}}ds\mathcal{L}\left( \theta ,%
\dot{\theta}\right)
\end{equation}%
where $\dot{\theta}=\frac{d\theta }{ds}$ and $\mathcal{L}(\theta ,\dot{\theta%
})$ is the Lagrangian of the system,%
\begin{equation}
\mathcal{L}(\theta ,\dot{\theta})=(g_{ij}\dot{\theta}^{i}\dot{\theta}^{j})^{%
\frac{1}{2}}\text{.}
\end{equation}%
The evolution of the system can be deduced from a variational principle of
the Jacobi type. A convenient choice for the affine parameter \ $s$ is one
satisfying the condition $g_{ij}\frac{d\theta ^{i}}{d\tau }\frac{d\theta ^{j}%
}{d\tau }=1$. Therefore we formally identify $s$ with the temporal evolution
parameter $\tau $. Performing a standard calculus of variations, we obtain,%
\begin{equation}
\delta J_{RED}\left[ \theta \right] =\int d\tau \left( \frac{1}{2}\frac{%
\partial g_{ij}}{\partial \theta ^{k}}\dot{\theta}^{i}\dot{\theta}^{j}-\frac{%
d\dot{\theta}_{k}}{d\tau }\right) \delta \theta ^{k}=0,\forall \delta \theta
^{k}\text{.}
\end{equation}%
Note that from $\left( 18\right) $, $\frac{d\dot{\theta}_{k}}{d\tau }=\frac{1%
}{2}\frac{\partial g_{ij}}{\partial \theta ^{k}}\dot{\theta}^{i}\dot{\theta}%
^{j}$. This "equation of motion" is interesting because it shows that if $%
\frac{\partial g_{ij}}{\partial \theta ^{k}}=0$ for a particular $k$ then
the corresponding $\dot{\theta}_{k}$ is conserved. This suggests to
interpret $\dot{\theta}_{k}$ as momenta. Equations $(18)$ and $\left(
11\right) $ lead to the geodesic equations,%
\begin{equation}
\frac{d^{2}\theta ^{k}(\tau )}{d\tau ^{2}}+\Gamma _{ij}^{k}\frac{d\theta
^{i}(\tau )}{d\tau }\frac{d\theta ^{j}(\tau )}{d\tau }=0\text{.}
\end{equation}%
Observe that $\left( 19\right) $ are second order equations. These equations
describe a dynamics that is reversible and they give the trajectory between
an initial and final position. The trajectory can be equally well traversed
in both directions.

\subsection{Geodesics on $\mathcal{M}_{s}$}

We seek the explicit form of $(19)$ for the statistical coordinates $(\mu
_{1},\mu _{2},\sigma _{2})$ parametrizing the submanifold $\mathit{m}_{s}$
of $\mathcal{M}_{s}$, $\mathit{m}_{s}=\left\{ p^{(tot)}\left( \vec{x}|\vec{%
\theta}\right) \in \mathcal{M}_{s}:\vec{\theta}\text{ satisfies }\left(
19\right) \right\} $. Substituting the explicit expression of the connection
coefficients found in subsection $(2.3)$ into $\left( 19\right) $, the
geodesic equations become,%
\begin{equation}
\begin{array}{c}
\frac{d^{2}\mu _{1}}{d\tau ^{2}}-\frac{1}{\mu _{1}}\left( \frac{d\mu _{1}}{%
d\tau }\right) ^{2}=0\text{, }\frac{d^{2}\mu _{2}}{d\tau ^{2}}-\frac{2}{%
\sigma _{2}}\frac{d\mu _{2}}{d\tau }\frac{d\sigma _{2}}{d\tau }=0\text{,} \\ 
\\ 
\frac{d^{2}\sigma _{2}}{d\tau ^{2}}-\frac{1}{\sigma _{2}}\left( \frac{%
d\sigma _{2}}{d\tau }\right) ^{2}+\frac{1}{2\sigma _{2}}\left( \frac{d\mu
_{2}}{d\tau }\right) ^{2}=0\text{.}%
\end{array}%
\end{equation}%
This is a set of coupled ordinary differential equations, whose solutions
have been obtained by use of mathematics software (Maple) and analytical
manipulation:%
\begin{equation}
\begin{array}{c}
\mu _{1}\left( \tau \right) =A_{1}\left( \cosh \left( \alpha _{1}\tau
\right) -\sinh \left( \alpha _{1}\tau \right) \right) \text{,} \\ 
\\ 
\mu _{2}\left( \tau \right) =\dfrac{A_{2}^{2}}{2\alpha _{2}}\dfrac{1}{\cosh
\left( 2\alpha _{2}\tau \right) -\sinh \left( 2\alpha _{2}\tau \right) +%
\frac{A_{2}^{2}}{8\alpha _{2}^{2}}}+B_{2}\text{,} \\ 
\\ 
\text{ }\sigma _{2}\left( \tau \right) =A_{2}\dfrac{\cosh \left( \alpha
_{2}\tau \right) -\sinh \left( \alpha _{2}\tau \right) }{\cosh \left(
2\alpha _{2}\tau \right) -\sinh \left( 2\alpha _{2}\tau \right) +\frac{%
A_{2}^{2}}{8\alpha _{2}^{2}}}\text{.}%
\end{array}%
\end{equation}%
The quantities $A_{1}$, $A_{2}$, $B_{2}$, $\alpha _{1}$ and $\alpha _{2}$
are the five integration constants ($5=6-1$, $\left( \dot{\theta}_{j}\dot{%
\theta}^{j}\right) ^{\frac{1}{2}}=1$). The coupling between the parameters $%
\mu _{2}$ and $\sigma _{2}$ is reflected by the fact that their respective
evolution equations in $(21)$ are defined in terms of the same integration
constants $A_{2}$ and $\alpha _{2}$. Equations $(21)$ parametrize the
evolution surface of the statistical submanifold $\mathit{m}_{s}\subset 
\mathcal{M}_{s}$. By eliminating the parameter $\tau $, $\sigma _{2}$ can be
expressed explicitly as a function of $\mu _{1}$ and $\mu _{2}$,%
\begin{equation}
\sigma _{2}\left( \mu _{1}\text{, }\mu _{2}\right) =\frac{2\alpha _{2}}{%
A_{1}^{\frac{\alpha _{2}}{\alpha _{1}}}A_{2}}\mu _{1}^{\frac{\alpha _{2}}{%
\alpha _{1}}}\left( \mu _{2}-B_{2}\right) \text{.}
\end{equation}%
This equation describes the submanifold evolution surface. To give a
qualitative sense of this surface, we plot $\left( 22\right) $ in Figure 1
for a special choice of a $1d$ set of initial conditions ($\alpha
_{2}=2\alpha _{1}$ while $A_{1}$, $A_{2}$ and $B_{2}$ are arbitrary).
Equations $\left( 20\right) $ are used to evolve this $1d$ line to generate
the $2d$ surface of $\mathit{m}_{s}$. This figure is indicative of the
instability of geodesics under small perturbations of initial conditions.%

\begin{figure}[!t]
 \resizebox{1.0\columnwidth}{!}
  {\includegraphics[draft=false]{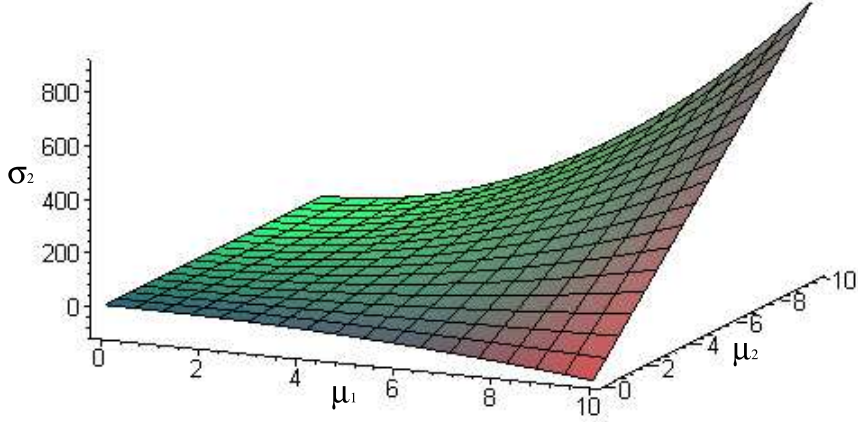}}
  \caption{The Statistical Submanifold Evolution Surface}
 \label{fig:a}
\end{figure}


\section{About the Stability of Geodesics on $\mathcal{M}_{s}$}

We briefly investigate the stability of the trajectories of the RED model
considered on $\mathcal{M}_{s}$. It is known $\left[ 8\right] $ that the
Riemannian curvature of a manifold is closely connected with the behavior of
the geodesics on it. If the Riemannian curvature of a manifold is negative,
geodesics (initially parallel) rapidly diverge from one another. For the
sake of simplicity, we assume very special initial conditions: $\alpha
=\alpha _{1}=\alpha _{2}\ll \frac{1}{4}$, $\frac{A_{2}}{8\alpha _{2}^{2}}\ll
1$; $A_{1}$ and $B_{2}$ are arbitrary. However, the conclusion we reach can
be generalized to more arbitrary initial conditions. Recall that $\mathcal{M}%
_{s}$ is the space of probability distributions $p^{(tot)}\left( \vec{x}|%
\vec{\theta}\right) $ labeled by parameters $\mu _{1},$ $\mu _{2},\sigma
_{2} $. These parameters are the coordinates for the point $p^{(tot)}$, and
in these coordinates a volume element $dV_{\mathcal{M}_{s}}$ reads, 
\begin{equation}
dV_{\mathcal{M}_{s}}=g^{\frac{1}{2}}\left( \vec{\theta}\right) d^{3}\vec{%
\theta}\equiv \sqrt{g}d\mu _{1}d\mu _{2}d\sigma _{2}
\end{equation}%
where $g=|\det \left( g_{ij}\right) |=\frac{2}{\mu _{1}^{2}\sigma _{2}^{4}}$%
. Hence, using $\left( 23\right) $, the volume of an extended region $\Delta
V_{\mathcal{M}_{s}}$ of $\mathcal{M}_{s}$ is,%
\begin{equation}
\Delta V_{\mathcal{M}_{s}}\left( \tau ;\alpha \right) =V_{\mathcal{M}%
_{s}}\left( \tau \right) -V_{\mathcal{M}_{s}}\left( 0\right)
=\dint\limits_{\mu _{1}\left( 0\right) }^{\mu _{1}\left( \tau \right)
}\dint\limits_{\mu _{2}\left( 0\right) }^{\mu _{2}\left( \tau \right)
}\dint\limits_{\sigma _{2}\left( 0\right) }^{\sigma _{2}\left( \tau \right) }%
\sqrt{g}d\mu _{1}d\mu _{2}d\sigma _{2}\text{.}
\end{equation}%
Finally, using $\left( 21\right) $ in $\left( 24\right) $, the temporal
evolution of the volume $\Delta V_{\mathcal{M}_{s}}$ becomes,%
\begin{equation}
\Delta V_{\mathcal{M}_{s}}\left( \tau ;\alpha \right) =\frac{A_{2}\tau }{%
\sqrt{2}}e^{\alpha \tau }.
\end{equation}%
Equation $\left( 25\right) $ shows that volumes $\Delta V_{\mathcal{M}%
_{s}}\left( \tau ;\alpha \right) $ increase exponentially with $\tau $.
Consider the one-parameter $\left( \alpha \right) $ family of statistical
volume elements $\mathcal{F}_{V_{\mathcal{M}_{s}}}\left( \alpha \right)
\equiv \left\{ \Delta V_{\mathcal{M}_{s}}\left( \tau ;\alpha \right)
\right\} _{\alpha }$. Note that $\alpha \equiv \alpha _{1}=$ $-\left( \frac{1%
}{\mu _{1}}\frac{d\mu _{1}}{d\tau }\right) _{\tau =0}>0$. The stability of
the geodesics of the RED model may be studied from the behavior of the ratio 
$\mathit{r}_{V_{\mathcal{M}_{s}}}$ of neighboring volumes $\Delta V_{%
\mathcal{M}_{s}}\left( \tau ;\alpha +\delta \alpha \right) $ and $\Delta V_{%
\mathcal{M}_{s}}\left( \tau ;\alpha \right) $,%
\begin{equation}
\mathit{r}_{V_{\mathcal{M}_{s}}}\overset{\text{def}}{=}\frac{\Delta V_{%
\mathcal{M}_{s}}\left( \tau ;\alpha +\delta \alpha \right) }{\Delta V_{%
\mathcal{M}_{s}}\left( \tau ;\alpha \right) }\text{.}
\end{equation}%
Positive $\delta \alpha $ is considered. The quantity $\mathit{r}_{V_{%
\mathcal{M}_{s}}}$ describes the relative volume changes in $\tau $ for
volume elements with parameters $\alpha $ and $\alpha +\delta \alpha $.
Substituting $\left( 25\right) $ in $\left( 26\right) $, we obtain%
\begin{equation}
\mathit{r}_{V_{\mathcal{M}_{s}}}=e^{\delta \alpha \cdot \tau }\text{.}
\end{equation}%
Equation $\left( 27\right) $ shows that the relative volume change ratio
diverges exponentially under small perturbations of the initial conditions.
Another useful quantity that encodes relevant information about the
stability of neighbouring volume elements might be the \textit{entropy-like}
quantity $S$ defined as,%
\begin{equation}
S\overset{\text{def}}{=}\log V_{\mathcal{M}_{s}}
\end{equation}%
where $V_{\mathcal{M}_{s}}$ is the average statistical volume element
defined as,%
\begin{equation}
V_{\mathcal{M}_{s}}\equiv \left\langle \Delta V_{\mathcal{M}%
_{s}}\right\rangle _{\tau }\overset{\text{def}}{=}\frac{1}{\tau }%
\dint\limits_{0}^{\tau }\Delta V_{\mathcal{M}_{s}}\left( \tau ^{\prime
};\alpha \right) d\tau ^{\prime }\text{.}
\end{equation}%
Indeed, substituting $\left( 25\right) $ in $\left( 29\right) $, the
asymptotic limit of $\left( 28\right) $ becomes,%
\begin{equation}
S\approx \alpha \tau \text{.}
\end{equation}%
Doesn't equation $\left( 30\right) $ resemble the Zurek-Paz chaos criterion $%
\left[ 9\text{, }10\right] $ of linear entropy increase under stochastic
perturbations? This question and a detailed investigation of the instability
of neighbouring geodesics on different curved statistical manifolds are
addressed in $\left[ 12\right] $ by studying the temporal behaviour of the
Jacobi field intensity $\left[ 11\right] $ on such manifolds.

Our considerations suggest that suitable RED models may be constructed to
describe chaotic dynamical systems and, furthermore, that a more careful
analysis may lead to the clarification of the role of curvature in inferent
methods for physics $\left[ 12\text{, }13\right] $.

\section{Final Remarks}

A RED model is considered. The space of microstates is $2D$ while all
information necessary to study the dynamical evolution of such a system is
contained in a $3D$ space of macrostates $\mathcal{M}_{s}$. It was shown
that $\mathcal{M}_{s}$ possess the geometry of a curved manifold of constant
negative curvature $\left( R=-1\right) $. The geodesics of the RED model are
hyperbolic curves on the submanifold $\mathit{m}_{s}$ of $\mathcal{M}_{s}$.
Furthermore, considerations of statistical volume elements suggest that
these entropic dynamical models might be useful to mimic exponentially
unstable systems. Provided the correct variables describing the true degrees
of freedom of a system be identified, ED may lead to insights into the
foundations of models of physics.

\noindent \textbf{Acknowledgements:} The authors are grateful to Prof. Ariel Caticha for very useful comments.

\end{document}